**Title**

Kernel Approximate Bayesian Computation in Population Genetic Inferences

**Authors**

Shigeki Nakagome[1], Kenji Fukumizu[1], Shuhei Mano[1,2,*]

**Affiliations**

[1]Department of Mathematical Analysis and Statistical Inference, The Institute of Statistical Mathematics, 10-3 Midori-cho, Tachikawa, Tokyo 190-8562, Japan

[2]Japan Science and Technology Agency, 4-1-8, Honcho, Kawaguchi-shi, Saitama 332-0012, Japan

[*]**Corresponding Author**

Shuhei Mano

Department of Mathematical Analysis and Statistical Inference, The Institute of Statistical Mathematics, 10-3 Midori-cho, Tachikawa, Tokyo 190-8562, Japan

Tel/Fax: +81-550-5533-8432/+81-42-526-4339

E-mail address: smano@ism.ac.jp

**Abstract**

Approximate Bayesian computation (ABC) is a likelihood-free approach for Bayesian inferences based on a rejection algorithm method that applies a tolerance of dissimilarity between summary statistics from observed and simulated data. Although several improvements to the algorithm have been proposed, none of these improvements avoid the following two sources of approximation: 1) lack of sufficient statistics: sampling is not from the true posterior density given data but from an approximate posterior density given summary statistics; and 2) non-zero tolerance: sampling from the posterior density given summary statistics is achieved only in the limit of zero tolerance. The first source of approximation can be improved by adding a summary statistic, but an increase in the number of summary statistics could introduce additional variance caused by the low acceptance rate. Consequently, many researchers have attempted to develop techniques to choose informative summary statistics. The present study evaluated the utility of a kernel-based ABC method (Fukumizu et al. 2010, arXiv:1009.5736 and 2011, NIPS 24: 1549-1557) for complex problems that demand many summary statistics. Specifically, kernel ABC was applied to population genetic inference. We demonstrate that, in contrast to conventional ABCs, kernel ABC can incorporate a large number of summary statistics while maintaining high performance of the inference.

# 1. Introduction

Approximate Bayesian computation (ABC) (Fu and Li 1997; Tavaré et al. 1997; Weiss and von Haeseler 1998; Pritchard et al. 1999; Beaumont et al. 2002) is a popular method to obtain an approximation of the posterior estimates without evaluating the likelihood. Assume observed discrete data, $\mathcal{D}$, are generated by a model that has parameters of interest, $\boldsymbol{\theta}$, that are generated by the prior density, $\pi(\boldsymbol{\theta})$. By Bayes' rule, the posterior density of $\boldsymbol{\theta}$ given $\mathcal{D}$ is $\pi(\boldsymbol{\theta}|\mathcal{D}) \propto f(\mathcal{D}|\boldsymbol{\theta})\pi(\boldsymbol{\theta})$, where $f(\mathcal{D}|\boldsymbol{\theta})$ is the likelihood of the model. Rejection sampling is a basic algorithm for sampling parameters from the posterior density. The algorithm takes the following form:

**Algorithm A.**

A1. Generate $\boldsymbol{\theta}'$ from $\pi(\cdot)$.

A2. Accept $\boldsymbol{\theta}'$ with probability proportional to $f(\mathcal{D}|\boldsymbol{\theta}')$, and go to A1.

Even when the likelihood is unknown, it is possible to sample parameters from the posterior density assuming data can be simulated under the model. In this case, A2 is replaced with the following:

A2′. Simulate data, $\mathcal{D}'$, by the model using $\boldsymbol{\theta}'$.

A3′. Accept $\boldsymbol{\theta}'$ if $\mathcal{D}=\mathcal{D}'$, and go to A1.

Although this algorithm gives samples from the true posterior density, the acceptance rate decreases sharply with increasing dimensionality of the data. Therefore, we introduce summary statistics for $\mathcal{D}$, which are denoted by $\boldsymbol{s}$, and step A3′ is replaced with A3″, as follows: accept $\boldsymbol{\theta}'$ if $d(\boldsymbol{s}, \boldsymbol{s}') < \eta$, where $d$ is a metric that measures the dissimilarity, $\boldsymbol{s}'$ gives the summary statistics of $\mathcal{D}'$, and $\eta$ is the tolerance (Fu and Li 1997). The algorithm involving rejection sampling with A2′ and A3″ corresponds to basic ABC and is specified hereafter as rejection ABC.

Although several improvements to the algorithm have been proposed, no improvements are capable of avoiding the following two sources of approximation: 1) lack of sufficient statistics: sampling is not from the true posterior density given data, $\pi(\boldsymbol{\theta}|\mathcal{D})$, but from an approximate posterior density given summary statistics, $\pi(\boldsymbol{\theta}|\boldsymbol{s})$; and 2) the non-zero tolerance: sampling from the posterior density, $\pi(\boldsymbol{\theta}|\boldsymbol{s})$, is achieved only in the limit $\eta \to 0$, but the acceptance rate decreases with decreasing $\eta$.

If the set of summary statistics is sufficient, we can sample from the true posterior density given data, $\pi(\boldsymbol{\theta}|\mathcal{D})$. However, it generally is difficult to obtain a set of sufficient statistics for a complex data set. A partition of a set, $\mathcal{D}$, is a set of nonempty subsets of $\mathcal{D}$ such that every element in $\mathcal{D}$ is in exactly one of these subsets. Then, a partition, $T$, of a set, $\mathcal{D}$, is called a refinement of a partition $S$ of $\mathcal{D}$ if every element of $T$ is a subset of some element of $S$. In population genetics, for example, the site frequency spectrum (SFS) gives a refinement of a partition of data that is determined by conventional summary statistics including the number of segregating sites, the nucleotide diversity (Nei and Li 1979), and Tajima's D (Tajima 1989). We may improve the approximation of the true posterior density given data by adding a summary statistic, which introduces further refinement of the partition of a data. For example, for independently and identically distributed data $\boldsymbol{x}$ from location density $f(x - \theta)$, the median of $\boldsymbol{x}$ is not a sufficient statistic for $\theta$, but the set of order statistics of $\boldsymbol{x}$ is sufficient (for example, Casella and Berger 2002). Unfortunately, there is no general rule to construct a set of summary statistics that closely approximates the posterior density. Refinement does not always improve the approximation. However, we note the following observation:

**Proposition**. *Suppose a set of summary statistics, **T**, determines a refinement of a partition of a data $\mathcal{D}$ that itself is determined by a set of summary statistics, **S**. Then, $\pi(\boldsymbol{\theta}|\boldsymbol{T}=\boldsymbol{t})$ gives a better approximation of $\pi(\boldsymbol{\theta}|\mathcal{D})$ than $\pi(\boldsymbol{\theta}|\boldsymbol{S}=\boldsymbol{s})$ in terms of the Kullback-Leibler divergence if the expectation of $\pi(\boldsymbol{\theta}|\boldsymbol{S}=\boldsymbol{s})/\pi(\boldsymbol{\theta}|\boldsymbol{T}=\boldsymbol{t})$ with respect to $\pi(\boldsymbol{\theta}|\mathcal{D})$ is smaller than unity.*

**Proof**: We have $D_{KL}(\pi(\boldsymbol{\theta}|\mathcal{D})|\pi(\boldsymbol{\theta}|\boldsymbol{s})) - D_{KL}(\pi(\boldsymbol{\theta}|\mathcal{D})|\pi(\boldsymbol{\theta}|\boldsymbol{t})) = \int \pi(\boldsymbol{\theta}|\mathcal{D}) \log \frac{\pi(\boldsymbol{\theta}|\boldsymbol{t})}{\pi(\boldsymbol{\theta}|\boldsymbol{s})} d\boldsymbol{\theta}$.

Using Jensen's inequality, we have $\int \pi(\boldsymbol{\theta}|\mathcal{D}) \log \frac{\pi(\boldsymbol{\theta}|\boldsymbol{t})}{\pi(\boldsymbol{\theta}|\boldsymbol{s})} d\boldsymbol{\theta} \geq -\log \mathrm{E}\left[\frac{\pi(\boldsymbol{\theta}|\boldsymbol{s})}{\pi(\boldsymbol{\theta}|\boldsymbol{t})}\Big|\mathcal{D}\right]$. Therefore, $D_{KL}(\pi(\boldsymbol{\theta}|\mathcal{D})|\pi(\boldsymbol{\theta}|\boldsymbol{s})) \geq D_{KL}(\pi(\boldsymbol{\theta}|\mathcal{D})|\pi(\boldsymbol{\theta}|\boldsymbol{t}))$ if $\mathrm{E}\left[\frac{\pi(\boldsymbol{\theta}|\boldsymbol{s})}{\pi(\boldsymbol{\theta}|\boldsymbol{t})}\Big|\mathcal{D}\right] \leq 1$, namely, the expectation of $\pi(\boldsymbol{\theta}|\boldsymbol{S}=\boldsymbol{s})/\pi(\boldsymbol{\theta}|\boldsymbol{T}=\boldsymbol{t})$ with respect to $\pi(\boldsymbol{\theta}|\mathcal{D})$ is smaller than unity. ∎

Intuitively, this proposition says that the Kullback-Leibler divergence decreases if the density is further conditioned by a set of summary statistics that lends more probability mass around the mode of the true posterior density given data. A simple example which is given in the supplementary materials is helpful to understand this situation.

An important issue interferes with attempts to improve the approximation of the true posterior density by increasing the number of summary statistics: a trade-off exists between information added by a new statistic and additional variance caused by the low acceptance rate. Consequently, many investigators have attempted to develop techniques to choose informative summary statistics (Joyce and Marjoram 2008; Wegmann et al. 2009; Blum and Francois 2010; Nunes and Balding 2010; Fearnhead and Prangle 2012). The selection of informative summary statistics is not straightforward. In an exhaustive search, a huge number of combinations must be assessed, whereas in a greedy search, the final set depends on the order in which the statistics are tested for inclusion. Nunes and Balding (2010) reported that the optimal set of summary statistics was specific to each data set and could not be generalized. Therefore, algorithms that avoid dimensional reduction are desirable.

The choice of the positive tolerance involves a trade-off between the bias and the variance such that increasing the tolerance reduces the variance by allowing a large number of accepted samples while simultaneously increasing the bias arising from the prior values. Several studies have addressed the approximation introduced by non-zero tolerance (Sisson et al. 2007; Beaumont et al. 2009; Drovandi and Pettit 2011) by proposing schemes to successively reduce tolerance with the sequential Monte Carlo algorithm. Consistency of the estimator is only attained in the limit of zero tolerance at the expense of large variances owing to poor acceptance rates. Fearnhead and Prangle (2012) introduced noisy-ABC, which calibrates the estimator, $P(\boldsymbol{\theta} \in \boldsymbol{\Theta} | P_\eta(\boldsymbol{\theta} \in \boldsymbol{\Theta} | \boldsymbol{s}) = p, \mathcal{D}) = p$ for $\forall \boldsymbol{\Theta}$, where $P_\eta(\boldsymbol{\theta} \in \boldsymbol{\Theta} | \boldsymbol{s})$ is the probability assigned by the posterior density obtained by ABC with tolerance $\eta$. Noisy-ABC is free from bias with the expense of introducing additional variance.

Kernel methods provide systematic data analysis by mapping variables into a reproducing kernel Hilbert space (RKHS) to extract nonlinearity or higher-order moments of data (Hofmann et al. 2008). An advantage of kernel methods is their computational efficiency when processing high-dimensional data. Whereas kernel methods employ high-dimensional nonlinear mappings from the data space to the RKHS, the inner products among data points are computed only with positive definite kernels instead of using the explicit form of the high-dimensional mapping. Thus, computing costs do not increase with the dimensionality of the data (*i.e.*, the number of summary statistics in ABC) but with the number of observations (*i.e.*, the number of simulations in ABC). The mean of mappings into the RKHS (*i.e.*, the kernel mean) recently was proposed to represent a probability distribution that could be applied to various data analyses (Smola et al. 2007). Fukumizu et al. (2010, 2011) applied this research to develop kernel methods of implementing Bayes' rule that enable one to compute the kernel mean representation of the posterior distribution in the form of a weighted sum of the sampled values. In addition, Fukumizu et al. (2010, 2011) proposed a new ABC approach, hereafter described as kernel ABC, and demonstrated its performance using a toy problem.

The present study evaluated the utility of kernel ABC for actual complex problems that demand a large number of summary statistics. Specifically, we demonstrate an application of the kernel ABC method to population genetic inferences. In contrast to conventional ABCs, kernel ABC can incorporate a large number of summary statistics while maintaining consistency of the posterior estimates and without compromising the performance of the inference.

## 2. Methods

Suppose the linear regression of parameters into summary statistics. The estimator of the slope, $\boldsymbol{\beta}$, minimizes $\sum_{i=1}^{n}\|\boldsymbol{\theta}_i - \langle \boldsymbol{\beta}, s_i \rangle\|^2$, where $\langle \cdot, \cdot \rangle$ and $\|\cdot\|$ denote the inner product and the norm in the Euclidian space, respectively. We may regard $\langle \widehat{\boldsymbol{\beta}}, s_i \rangle$ as an estimator of the conditional mean because the conditional mean minimizes the mean squared error (MSE) pointwise as follows: $E[\boldsymbol{\theta}|\boldsymbol{S} = \boldsymbol{s}] = \mathrm{argmin}_c E_{\boldsymbol{\theta}|\boldsymbol{s}}\left(\|\boldsymbol{\theta} - c\|^2 | \boldsymbol{S} = \boldsymbol{s}\right)$ (Hastie, Tibshirani, and Friedman, 2009). Kernel ABC is introduced by a simple extension of this procedure using a kernel-based method.

Consider a map $\Phi: \Omega \to \mathcal{H}_S$ defined by $\Phi(\boldsymbol{s}) = k(\cdot, \boldsymbol{s})$, where $\Omega$ is a space of summary statistics, and $\mathcal{H}_S$ is the RKHS associated with a positive definite kernel, $k$. The most useful property of the RKHS is the reproducing property. The function value is given by the inner product as $\langle f(\cdot), k(\cdot, \boldsymbol{s}_i)\rangle_{\mathcal{H}_S} = f(\boldsymbol{s}_i)$ for $\forall f \in \mathcal{H}_S$, where $\langle \cdot, \cdot \rangle_{\mathcal{H}_S}$ is the inner product in $\mathcal{H}_S$. In kernel methods, data, $\boldsymbol{s}$, are mapped into the RKHS as $\Phi(\boldsymbol{s}) = k(\cdot, \boldsymbol{s})$, and $\Phi(\boldsymbol{s})$ is regarded as a feature vector of $\boldsymbol{s}$. The inner product between summary statistic values $\boldsymbol{s}$ and $\boldsymbol{s}_i$ thus is given by $\langle \Phi(\boldsymbol{s}), \Phi(\boldsymbol{s}_i)\rangle_{\mathcal{H}_S} = k(\boldsymbol{s}, \boldsymbol{s}_i)$, which is called a kernel trick and is the basis of the efficient computation of kernel methods. In kernel methods, a probability distribution of $s$ is expressed by the mean $E[k(\cdot, s)]$ of the random feature vector $k(\cdot, s)$ in the RKHS, which is known to be sufficient to determine the distribution uniquely with an appropriate choice of kernel. The distribution of $\boldsymbol{s}$ is expressed by the mean of the random feature vector, $k(\cdot, \boldsymbol{s})$, in the RKHS, which is called the kernel mean.

The empirical estimator of the kernel posterior mean operator of $\boldsymbol{\theta}$ given an observation, $\boldsymbol{s}$, by $n$ simulations, $\{(\boldsymbol{\theta}_i, \boldsymbol{s}_i)\}_{i=1}^n$, is given by $\hat{\mu}_{\boldsymbol{\theta}|\boldsymbol{s}} = \sum_{i=1}^n w_i\, k(\cdot, \boldsymbol{\theta}_i)$ (Song et al. 2009; Fukumizu et al. 2010), where $\boldsymbol{\theta}_i$ is a set of parameters generated by the $i$-th simulation. Taking the inner product with a function in the RKHS, the operator gives an estimator of the conditional mean of the function. The weight, $w_i$, is given by $w_i = \sum_{j=1}^n (G_S + n\varepsilon_n I_n)^{-1}_{ij}\, k(\boldsymbol{s}_j, \boldsymbol{s})$, where $G_S$ is the Gram matrix consisting of $\left(k(\boldsymbol{s}_i, \boldsymbol{s}_j)\right)_{i,j=1}^n$, $\varepsilon_n$ is the coefficient of the Tikhonov-type regularization that biases the data to stabilize the matrix inversion, and $I_n$ is the identity matrix.

The estimator can be obtained by systematic construction (Fukumizu et al. 2010, 2011), but it follows immediately from the kernel ridge regression of the parameter onto the summary statistics. Consider the estimate of the slope, $\boldsymbol{\beta} \in \mathcal{H}_S$, by minimizing

$$\sum_{i=1}^n \left\| \boldsymbol{\theta}_i - \langle \boldsymbol{\beta}, \Phi(\boldsymbol{s}_i)\rangle_{\mathcal{H}_S} \right\|^2 + n\varepsilon_n \|\boldsymbol{\beta}\|^2. \qquad [1]$$

According to the representer theorem, the estimator of the posterior mean, $E[\boldsymbol{\theta}|\boldsymbol{s}]$, is given by $\langle \hat{\boldsymbol{\beta}}, \Phi(\boldsymbol{s})\rangle_{\mathcal{H}_S} = \sum_{i=1}^n w_i \boldsymbol{\theta}_i$. In the same manner, the posterior expectation of a function given by $f(\boldsymbol{\theta})$, $E[f(\boldsymbol{\theta})|\boldsymbol{s}]$ is estimated by $\langle f(\cdot), \hat{\mu}_{\boldsymbol{\theta}|\boldsymbol{s}}\rangle_{\mathcal{H}_S} = \sum_{i=1}^{n} w_i f(\boldsymbol{\theta}_i)$.

**Theorem (Song et al. 2009).** *The estimator $\hat{\mu}_{\boldsymbol{\theta}|\boldsymbol{s}}$ is consistent, that is*

$\langle f(\cdot), \hat{\mu}_{\boldsymbol{\theta}|\boldsymbol{s}}\rangle_{\mathcal{H}_S} - E[f(\boldsymbol{\theta})|\boldsymbol{s}] = O_p\left((n\varepsilon_n)^{-1/2} + \varepsilon_n^{1/2}\right)$ *as* $n \to \infty$.

**Proof.** A proof is given in the Appendix because Song et al. (2009) does not show a proof.

□

**Remark.** The first term corresponds to the variance, and the second term corresponds to bias introduced by the regularization.

 The kernel ridge regression adapted here is reasonable because we cannot avoid multicollinearity among the large number of summary statistics. An implementation of ABC, specified hereafter as regression ABC (Beaumont et al. 2002), uses a locally weighted regression with a smoothing kernel that is known to be weak for high-dimensional data (*i.e.*, more than several dimensions) (Loader 1999). In contrast, the kernel ridge regression achieves an error bound that does not explicitly depend on the dimensionality (Caponnetto and De Vito 2007) if the target function is in a certain class of smooth functions. Thus, kernel ridge regression is likely preferable for cases involving many summary statistics.

 The kernel ABC algorithm to compute the posterior expectation of a function, $f(\boldsymbol{\theta})$, takes the following form:

**Algorithm B (Fukumizu et al. 2010).**

B1. Generate $\boldsymbol{\theta}_i$ from $\pi(\boldsymbol{\theta})$.

B2. Simulate data, $\mathcal{D}_i$, by the model using $\boldsymbol{\theta}_i$.

B3. Compute the summary statistics, $\boldsymbol{s}_i$, for $\mathcal{D}_i$, and return to B1.

B4. Compute the estimator, $\sum_{i=1}^{n} w_i f(\boldsymbol{\theta}_i)$, with $\{(\boldsymbol{\theta}_i, \boldsymbol{s}_i)\}_{i=1}^{n}$.

Algorithm B is similar to the importance sampling, but the weights, $w_i$, are not positive-definite. Therefore, the estimator could give a nonsense result if the number of simulations is too small. For an estimator of the posterior density, B4 is replaced with B4′, as follows: Compute the estimator, $\sum_{i=1}^{n} w_i \delta_{\boldsymbol{\theta}_i}$, with $\{(\boldsymbol{\theta}_i, \boldsymbol{s}_i)\}_{i=1}^{n}$. By the argument on credible intervals in Section 3.1, we can probe that $\sum_{i=1}^{n} w_i \to 1$ as $n \to \infty$, which indicates that $\sum_{i=1}^{n} w_i \delta_{\boldsymbol{\theta}_i}$ is a signed measure.

A sharp contrast exists between the consistencies of the estimators using conventional ABCs versus kernel ABC. Conventional ABCs have consistency only in the limit $\eta \to 0$, whereas kernel ABC has consistency irrespective of the kernel choice. In conventional ABCs, it is possible to reduce bias by letting $\eta \to 0$ with $n \to 0$, but this compromises the acceptance. In kernel ABC, we can achieve consistency easily by scaling the regularization coefficient to the number of simulated samples as $\varepsilon_n \to 0$ such that $n\varepsilon_n \to \infty$ (see Theorem). Throughout this study, we scaled the coefficient as $\varepsilon_n \propto 1/\sqrt{n}$.

We implemented three conventional ABCs: rejection ABC, regression ABC (Beaumont et al. 2002), and a semi-automatic version of regression ABC (Fernhead & Prangles, 2012). Hereafter, we refer to the semi-automatic version of regression ABC as semi-automatic ABC. Summary statistics were standardized using the estimates of the means and the standard deviations obtained from 10,000 simulated datasets. For the implementation of kernel ABC, we used the Gaussian radial base function kernel, $k(x, y) = \exp(-\|x - y\|^2 / (2\sigma^2))$. The band width, $\sigma$, and the regularization parameter, $\varepsilon_n = a/\sqrt{n}$, were chosen by 10-fold cross validation by minimizing $\sum_{i=1}^{10} \left\| \frac{1}{|T_i|} \sum_{j \in T_i} \widehat{m}_{\boldsymbol{\theta}|\boldsymbol{s}_j}^{[-i]} - \widehat{m}_{\boldsymbol{\theta}}^{[i]} \right\|_{\mathcal{H}_{\boldsymbol{\theta}}}^{2}$, where $T_i$ represents a set of indices in the $i$-th subsample, $[i]$ represents the estimator based on the $i$-th subsample, and $[-i]$ represents the estimator based on all the subsamples except for the $i$-th subsample (Fukumizu et al. 2010, 2011).

In this study, we assessed the performance of inferences using an estimator of the MSE of the posterior mean, which was obtained by $l$ replications of construction of the estimator of the posterior mean according to algorithm B, $\frac{1}{l}\sum_{i=1}^{l} \left\| \widehat{m}_{i\boldsymbol{\theta}|\boldsymbol{s}} - m_{\boldsymbol{\theta}|\boldsymbol{s}} \right\|^2$, where $\widehat{m}_{i\boldsymbol{\theta}|\boldsymbol{s}}$ is the $i$-th estimate of the posterior mean, and $m_{\boldsymbol{\theta}|\boldsymbol{s}}$ is the true value of the posterior mean. According to the weak law of large numbers, the estimator tends to the MSE in probability as $l \to \infty$. Throughout this study, we set $l = 100$. For a case in which the likelihood was not available, the true value of the posterior mean was replaced by averaging 100 replications of the estimates obtained by kernel ABC with $n = 16{,}000$ samples. Here, we used kernel ABC to compute the true value because the posterior mean estimator obtained by kernel ABC is consistent (see Theorem) and should converge to the true value simply by using a large number of simulations (see Section 3.3).

Several reports have described the performance of inferences in terms of the sum of the squared error between the posterior estimates and the prior values (Beaumont et al. 2002; Wegmann et al. 2009; Nunes and Balding 2010). We considered the MSE because it is a simple interpretation of the sum of the variance and the squared bias of an estimator, and our primary interest in this study was to investigate how well the consistency is achieved in kernel ABC. In other words, a non-zero MSE as $n \to \infty$ is evidence of bias. If we have a consistent estimator, the MSE should decrease with increasing sample size because both the bias and the variance decrease.

## 3. Results

## 3.1 *Consistency*

We consider the constant-size population model to evaluate the consistency of the posterior mean. The constant-size model is advantageous because of its straightforward computation of likelihoods and because computations are implemented into the GENETREE software package (Griffiths and Tavaré 1994; Griffiths 2007). Therefore, we can sample from a true posterior density given data, $\pi(\theta|\mathcal{D})$, or from than given number of segregating sites ($S_{Seg}$), $\pi(\theta|S_{Seg})$ using simple rejection sampling.

We assumed a sample of 100 chromosomes taken from a population of constant size ($N = 10{,}000$) and a large (100 kb) nonrecombining region that was evolving under the infinite-sites mutation model (Kimura 1969; Watterson 1975). We regarded the population scaled mutation rate, $\theta = 4Mu$, as the parameter for which $M$ is the population size and $u$ is the fixed mutation rate per 100 kb per generation ($2.5 \times 10^{-4}$). Therefore, the true value of the parameter for the sample was $4Nu = 10$. We assumed a log-normal distribution for the prior density of $M$, which was described by a mean and a variance of $N$ and $N^2$, respectively. All simulations were conducted using the program package *ms*, which generates samples from the coalescent model (Hudson 2002).

The data, $\mathcal{D}$, were represented as numbers of sequences of several types, which were determined by the sequence of mutations experienced along the path to the most recent common ancestor of the sample. Data, $\mathcal{D}$, were summarized by the number of segregating sites, $S_{Seg}$, or the SFS. Because estimates of the SFS are unstable for samples consisting of 100 chromosomes, we used a coarse-grained spectrum consisting of 7 bins based on the Sturges' formula ($1 + \log_2 S_{Seg}$), which is denoted as $\boldsymbol{S}_{\text{SFS}}$. The frequencies were binned as follows: $0 - 8\%$, $8 - 16\%$, $16 - 24\%$, $24 - 32\%$, $32 - 40\%$, $40 - 48\%$, and $48 - 100\%$. We generated $10{,}000$ simulated datasets and calculated the average $\mathbf{S}_{\text{SFS}}$. We then chose a typical dataset as an observation that had the smallest sum of squared deviations from the average: $s_{Seg} = 49$ and $\mathbf{s}_{\text{SFS}} = (28,6,4,3,2,1,5)$. We used the near-average data set as a representative example. Unlikely data also were investigated, but the MSE did not change appreciably. This was expected because the MSE measures the discrepancy between the estimator of the posterior mean and the true posterior mean irrespective of how close the posterior mean is to the parameter values that were used to produce the dataset.

We computed the posterior mean given the number of segregating sites by rejection sampling simulations until 1 million samples were accepted. The estimate of the posterior mean, which was assumed to be the true value, was $m_{\theta|s_{Seg}} = 9.695$. We then estimated the MSE of the posterior mean estimator obtained by kernel ABC under different numbers of simulations: $n=1{,}000,\ 2{,}000,\ 4{,}000,\ 8{,}000,$ and $16{,}000$. The convergence of $\widehat{m}_{\theta|s_{Seg}}$ to the true value with increasing numbers of simulations is shown in Fig. 1. The consistency is readily confirmed by the fact that the MSE approaches zero, and the variance decreases with increasing numbers of simulations.

Next, we examined how kernel ABC accurately estimates the posterior credible interval given $S_{Seg}$. The consistency of the kernel estimator for credible intervals, $\sum_{i=1}^{n} w_i I_A(\boldsymbol{\theta}_i)$, can be proved by the convergence results for kernel ridge regression (Caponnetto and De Vito 2007), if we note that the estimator coincides with the solution to the regularized least square problem [1] with replacing $\boldsymbol{\theta}$ by $I_A(\boldsymbol{\theta})$. Here, $I_A(\cdot)$ is the indicator function of an interval $A$. We computed the 80% credible interval (from the 10th percentile to the 90th percentile) by performing rejection sampling simulations until 1 million samples were accepted. The estimates of the credible interval, which were assumed to be the true values, were 6.650–13.038. We then estimated the credible interval by kernel ABC under different numbers of simulations: $n=1,000$, 8,000, and 16,000. The estimates were 6.590–13.260, 6.548–13.021, and 6.675–13.113, for $n=1,000$, 8,000, and 16,000, respectively. These results demonstrate that the credible interval monotonically converges to the true interval.

3.2 *Improvement of approximation for high-dimensional summary statistics*

We computed the posterior mean given data, $\mathcal{D}$, using rejection sampling. The likelihood surface of $\theta$ was approximated by averaging the likelihoods at each point from 0.1 to 35.1 (bin width 1.0; total 35 points) over 0.1 billion simulations of GENETREE. We repeated the rejection sampling simulations using the likelihood surface until 1 million samples were accepted. The posterior mean of $\theta$ given $\mathcal{D}$, which was assumed to be the true value, was $m_{\theta|\mathcal{D}} = 10.498$.

Since the SFS determines a refinement of a partition given by the number of segregating sites, we can expect that the SFS improves the approximation of the true posterior mean given data compared with the number of segregating sites. The posterior mean given SFS, $\hat{m}_{\theta|s_{SFS}}$, was estimated to be 10.510 with a standard deviation of 0.044 using kernel ABC by averaging 100 replications with n=16,000 samples. We concluded that $S_{SFS}$ seems to improve $S_{Seg}$, as expected, because of the significant deviation of $m_{\theta|s_{Seg}} = 9.695$ from $m_{\theta|\mathcal{D}} = 10.498$, in contrast to $\hat{m}_{\theta|s_{SFS}} = 10.510$.

3.3 *Comparison with Conventional ABCs*

The performances of kernel ABC and conventional ABCs were evaluated in terms of the costs of computing times against fixed MSE values. Throughout this study, we implemented all computations in the C/C++ languages. Computations were conducted using an Intel Xeon X5680 3.33 GHz processor. Toward this end, we had to find the true posterior means of the parameters. The value of $m_{\theta|S_{Seg}} = 9.695$ had been generated by rejection sampling (See Section 3.1), and kernel ABC was used to estimate $\hat{m}_{\theta|S_{Seg}} = 9.686$ by averaging 100 replications with $n = 16{,}000$ samples. The similarity of these values suggested that averaging 100 replications of the estimates by kernel ABC with $n = 16{,}000$ samples was likely to give reliable estimates of the true value. In contrast, rejection ABC cannot achieve $\delta = 0$ with $S_{\text{SFS}}$; we found that no samples were accepted under $\eta = 0$ during a 2-week simulation. Therefore, we assumed that the true value of the posterior mean given $S_{\text{SFS}}$ was an average of 100 replications of the estimates obtained by kernel ABC with $n=16{,}000$ samples.

Estimates of the MSEs of $\hat{m}_{\theta|S_{Seg}}$ and $\hat{m}_{\theta|S_{\text{SFS}}}$ were calculated using differently sized simulations in kernel ABC ($n = 1{,}000,\ 2{,}000, 4{,}000,\ \text{and}\ 8{,}000$) and at different acceptance rates in three conventional ABC algorithms: rejection ABC, regression ABC, and semi-automatic ABC (for $\hat{m}_{\theta|S_{\text{SFS}}}$). Kernel ABC, regression ABC, and semi-automatic ABC are easy to compare because all of these algorithms are based on regression. The result for rejection ABC was included because it is the most basic ABC algorithm. For rejection ABC, tolerances were chosen such that 1,000 samples were accepted for each run of the simulation. The acceptance rate in regression ABC was defined as the proportion at which the Epanechnikov kernel gave a non-zero value (Beaumont, et al. 2002). We set a band width of the kernel and simulated samples until 1,000 samples were accepted. The results, displayed as computing times versus MSE estimates, are depicted with the standard deviations of the estimates in Fig. 2 and Fig. 3. When $S_{Seg}$ was used, the computational cost against a fixed MSE value was lower for the conventional ABCs than for the kernel ABC (Fig. 2). However, in contrast to the case with $S_{Seg}$, we observed that kernel ABC with $S_{SFS}$ significantly outperformed conventional ABCs in terms of computing times at a fixed MSE (Fig. 3). We found that regression ABC outperforms rejection ABC when $S_{SFS}$ is used. We also determined that semi-automatic ABC under-performs regression ABC, which suggests that constructing one-dimensional summary statistics may be over-summarizing. Our results suggest that kernel ABC gives better performance than conventional ABCs when higher-dimensional summary statistics are used.

3.4 *A Realistic Model*

In population genetics, the estimation of the scaled mutation rate is basic. As a more ambitious inference, we consider a more realistic model in which a population size bottleneck and subsequent expansion were assumed. We then tried to estimate the population sizes and the timings of the population size changes. We considered a sample of 100 chromosomes taken from a population and a large recombining region (100 kb) for which recombination was set at a fixed scaled rate of $\rho = 4N_3 u$. The assumed population demography was as follows: the ancestral size was $N_1 = 10,000$, and at time $T_2 = 4,000$ generations ago, the size instantaneously shrank to $N_2 = 2,000$. The size remained constant at size $N_2$ until $T_1 = 2,000$ generations ago, when it began expanding exponentially to reach size $N_3 = 20,000$ at present time. We regarded $\boldsymbol{\theta} = (M_1, M_2, M_3, U_1, U_2)$ as the parameters; the true values were $(N_1, N_2, N_3, T_1, T_2)$. We assumed a log-normal distribution for the prior density for the parameters. The means and variances of the parameters were the true values and the squared true values, respectively. We assumed $s_{\text{SFS}} = (16,1,1,1,1,1,4)$ for the observed data using the same procedure as the constant-size model. Since the SFS cannot account for recombination, we added a haplotype frequency spectrum (HFS) for summary statistics that consisted of haplotype frequencies in the sample. The SFS+HFS refines a partition of data given by SFS alone. The number of bins in the SFS and their intervals were identical to those in the constant-size model. The HFS was segregated into 7 bins as follows: $0-2\%$, $2-4\%$, $4-6\%$, $6-8\%$, $8-10\%$, $10-12\%$, $12-100\%$. For the SFS+HFS, we assumed $s_{\text{SFS+HFS}} = (18,0,0,0,1,0,1,13,2,1,0,0,0,2)$.

We compared the performances of kernel ABC, rejection ABC, regression ABC, and semi-automatic ABC. Although increasing the number of accepted samples recovered the performance of regression ABC, this algorithm demanded a huge computational time cost. In Fig. 4 and Fig. 5, computational times versus estimates are depicted with the standard deviation in which $s_{\text{SFS}}$ and $s_{\text{SFS+HFS}}$ are given. To scale the posterior means, we calculated the MSEs, which were standardized by the assumed true values. Specifically, the $i$-th estimate of the posterior mean, $\hat{m}_{i\boldsymbol{\theta}|\boldsymbol{s}}$, was standardized as $(\hat{m}_{i\boldsymbol{\theta}|\boldsymbol{s}} - m_{\boldsymbol{\theta}|\boldsymbol{s}})/m_{\boldsymbol{\theta}|\boldsymbol{s}}$, where $m_{\boldsymbol{\theta}|\boldsymbol{s}}$ was the true value of the posterior mean. Fig. 4 and Fig. 5 show that the computing time for kernel ABC was substantially lower for any fixed MSEs than the computing times using other conventional ABCs.

As reported previously (Beaumont et al. 2002), regression ABC did not exhibit a monotonic decrease in the MSE with decreasing tolerance. Beaumont et al. (2002) found that regression ABC eventually under-performed rejection ABC with decreasing tolerance.

This tendency is related to the fact that the decrease in band width of the Epanechnikov kernel in regression ABC increases the variance. If the band width is too narrow, the acceptance rate decreases. This tendency of regression ABC is related to the curse of dimensionality, by which many simulated summary statistics are closer to the boundary of the bands when a large number of summary statistics are used (Hastie et al. 2009). Choosing the optimal acceptance rate is the same as choosing the optimal band width of the Epanechnikov kernel.

## 4. Discussion

In this study methods are evaluated by how well we can approximate the posterior estimate given data. The use of large number of summary statistics is reasonable from the perspective of increasing model complexity, such as for population genetic analyses involving vast amounts of genomic data. To preserve tractable computation costs, conventional ABCs involve two sources of approximation: 1) lack of sufficient statistics and 2) non-zero tolerance to enrich acceptance rates. However, both of these aspects also influence the performance of the inference by introducing biases. Consequently, recent progress in ABC algorithm development has focused on techniques for selecting more informative summary statistics (Joyce and Marjoram 2008; Wegmann et al. 2009; Blum and Francois 2010; Nunes and Balding 2010; Fearnhead and Prangle 2012).

In this study, we pursued another direction that involved a kernel-based ABC method in which the dependence on non-informative summary statistics was automatically shrunken, thus avoiding the need to choose optimal set of summary statistics. We have demonstrated that kernel ABC successfully reduces the difficulties associated with conventional ABCs by applying ABCs to population genetic inferences. We found that kernel ABC is more efficient than conventional ABCs regarding computational times for fixed MSE values. We demonstrated that kernel ABC can accommodate a large number of summary statistics without compromising performance of the inference. With regard to tolerance, a posterior estimate obtained by kernel ABC maintained consistency. Therefore, the MSE of the estimators could be reduced simply by decreasing the variance and increasing the number of simulations. For an increased number of simulations, the low-rank matrix approximation (Fine and Scheinberg 2002) could substantially reduce the computation time involved in the Gram matrix inversion.

Choosing the band width and the regularization parameters by cross-validation is a computationally expensive step. We investigated a simpler alternative for choosing these parameters. For the band width, using the median of pairwise Euclidean distances in the simulated summary statistics generally worked well. For the regularization parameters, assessing the bias–variance trade-off in terms of the MSE gave similar values to those obtained by the cross-validation.

Recent progress in ABC algorithm development also has involved the use of Markov chain Monte Carlo or sequential Monte Carlo methods (Marjoram et al. 2003; Sisson et al. 2007; Beaumont et al. 2009; Wegmann et al. 2009; Drovandi and Pettit 2011). We did not address these approaches in the present study, but kernel methods could be useful to improve such methods.

**Acknowledgements**

We thank Mark Beaumont and Kevin Dawson for helpful discussions and comments.

We also thank anonymous referees for helpful comments and the proposal of Example in the text. S.N. was supported in part by a Grant-in-Aid for the Japan Society for the Promotion of Science (JSPS) Research fellow (24-3234). K.F. was supported in part by JSPS KAKENHI (B) 22300098.

8  **Appendix: Proof of Theorem**

9     This proof is similar to the proof of Theorem 6.1 in Fukumizu et al. (2010). Denote variance and covariance operators by $C_{ss}$ and $C_{\theta s}$, respectively, and the empirical estimators of them by $\hat{C}_{ss}$ and $\hat{C}_{\theta s}$, respectively. We have (Fukumizu et al. 2010, 2011)

$$\langle f(\cdot), \hat{\mu}_{\theta|s} \rangle_{\mathcal{H}_S} - E[f(\theta)|s] = \langle k(\cdot, \theta), \hat{C}_{\theta s}(\hat{C}_{ss} + \varepsilon_n I_n)^{-1} f(\cdot) - C_{\theta s} C_{ss}^{-1} f(\cdot) \rangle.$$

We want to establish

$$\left\| \hat{C}_{\theta s}(\hat{C}_{ss} + \varepsilon_n I_n)^{-1} f(\cdot) - C_{\theta s} C_{ss}^{-1} f(\cdot) \right\|_{\mathcal{H}_\theta} = O_p\left((n\varepsilon_n)^{-1/2} + \varepsilon_n^{1/2}\right). \qquad [2]$$

First we show

$$\left\| \hat{C}_{\theta s}(\hat{C}_{ss} + \varepsilon_n I_n)^{-1} f(\cdot) - C_{\theta s}(C_{ss} + \varepsilon_n I_n)^{-1} f(\cdot) \right\|_{\mathcal{H}_\theta} = O\left((n\varepsilon_n)^{-1/2}\right). \qquad [3]$$

The left hand side is upper bounded by

$$\left\| (\hat{C}_{\theta s} - C_{\theta s})(\hat{C}_{ss} + \varepsilon_n I_n)^{-1} f(\cdot) \right\|_{\mathcal{H}_\theta} +$$

$$\left\| \hat{C}_{\theta s}(\hat{C}_{ss} + \varepsilon_n I_n)^{-1} (\hat{C}_{ss} - C_{ss})(C_{ss} + \varepsilon_n I_n)^{-1} f(\cdot) \right\|_{\mathcal{H}_\theta}. \qquad [4]$$

By the decomposition $\hat{C}_{\theta s} = \hat{C}_{\theta\theta}^{1/2} \widehat{W}_{\theta s} \hat{C}_{ss}^{1/2}$ with $\|\widehat{W}_{\theta s}\| \leq 1$ (Baker 1973), we have

$$\left\| \hat{C}_{\theta s}(\hat{C}_{ss} + \varepsilon_n I_n)^{-1} \right\| = \left\| \hat{C}_{\theta\theta}^{1/2} \widehat{W}_{\theta s} \right\| \left\| \hat{C}_{ss}^{1/2}(\hat{C}_{ss} + \varepsilon_n I_n)^{-1/2} \right\| \left\| (\hat{C}_{ss} + \varepsilon_n I_n)^{-1/2} \right\| =$$

$O(\varepsilon_n^{-1/2})$. With the $\sqrt{n}$ consistency of the variance operator, we see that the second term of [4] is $O(\varepsilon_n^{-1/2})$. In a similar argument gives that the first term of [4] is $O((n\varepsilon_n)^{-1/2})$.

Therefore we have [3]. Then, we show

$$\left\|C_{\theta s}(C_{ss} + \varepsilon_n I_n)^{-1}f(\cdot) - C_{\theta s}C_{ss}^{-1}f(\cdot)\right\|_{\mathcal{H}_\theta} = O(\varepsilon_n^{1/2}). \qquad [5]$$

The left hand side is upper bounded by $\left\|C_{\theta\theta}^{1/2}W_{\theta s}\right\|\left\|C_{ss}^{1/2}(C_{ss} + \varepsilon_n I_n)^{-1}f(\cdot) - C_{ss}^{-1/2}f(\cdot)\right\|$.

By the eigendecomposition $C_{ss} = \sum_i \lambda_i \phi_i \langle \phi_i, \cdot \rangle$, where $\{\lambda_i\}$ are the eigenvalues and $\{\phi_i\}$ are the corresponding unit eigenvectors, we have a expansion

$$\left\|C_{ss}^{1/2}(C_{ss} + \varepsilon_n I_n)^{-1}f(\cdot) - C_{ss}^{-1/2}f(\cdot)\right\|_{\mathcal{H}_s}^2 = \sum_i \langle \phi_i, f \rangle^2 \left(\frac{\varepsilon_n \lambda_i^{1/2}}{\lambda_i + \varepsilon_n}\right)^2,$$

Where

$$\frac{\varepsilon_n \lambda_i^{1/2}}{\lambda_i + \varepsilon_n} = \frac{\lambda_i^{1/2}}{(\lambda_i + \varepsilon_n)^{1/2}} \frac{\varepsilon_n^{1/2}}{(\lambda_i + \varepsilon_n)^{1/2}} \varepsilon_n^{1/2} = O(\varepsilon_n^{1/2}).$$

Therefore we have [5]. Then, [2] follows from [3] and [5] ∎.

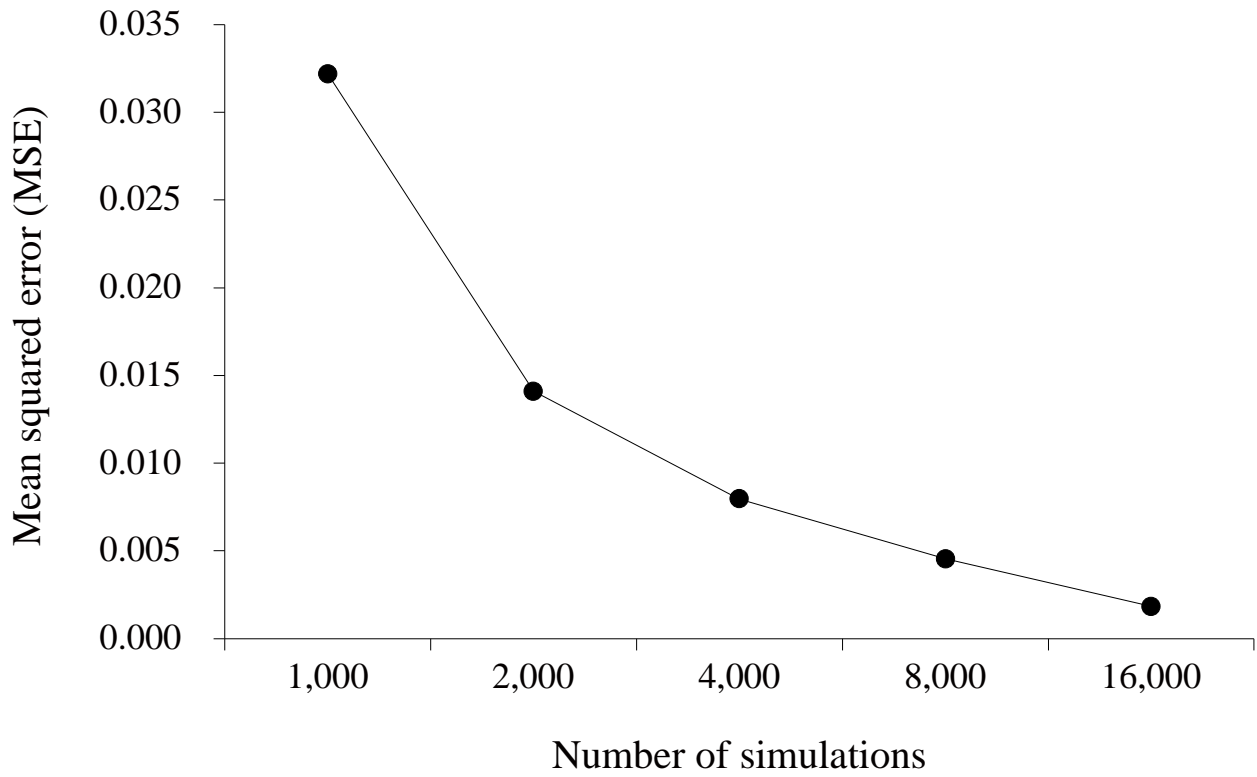

**Figure 1**. The rate of convergence of $\hat{m}_{\theta|S_{Seg}}$ on $m_{\theta|S_{Seg}}$. The x-axis indicates the number of simulations. The y-axis indicates MSE values.

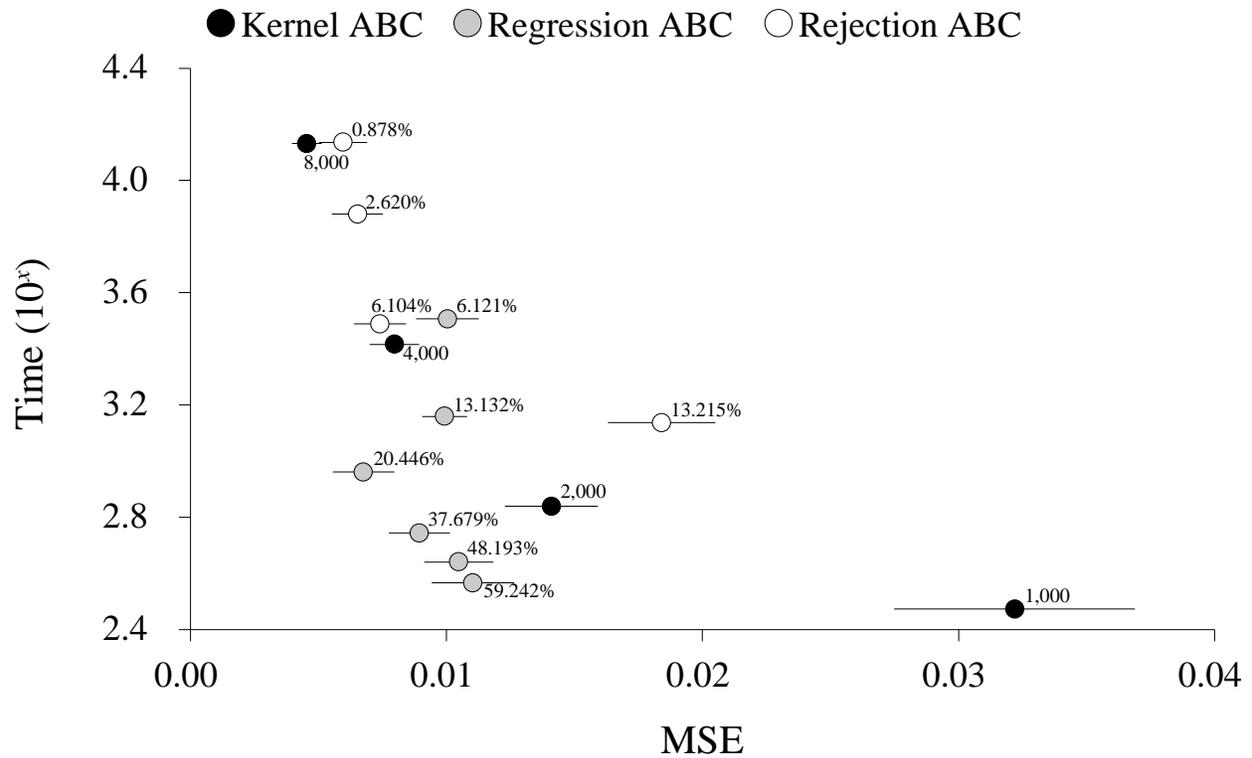

**Figure 2**. Comparisons of kernel-ABC and conventional ABCs under the constant size model using $s_{Seg}$. Computational costs scaled by $10^y$ seconds (y-axis) are plotted against MSE values (x-axis). Acceptance rates are also shown for rejection-ABC and regression-ABC. The tolerances were chosen so that 1,000 samples are accepted, so the number of simulations for each point is given by dividing 1,000 by the acceptance rates.

**Figure 3**. Comparisons of kernel-ABC and conventional ABCs under the constant size model using $s_{SFS}$. The inset compares the performances of kernel-ABC, regression-ABC, and semi-automatic-ABC.

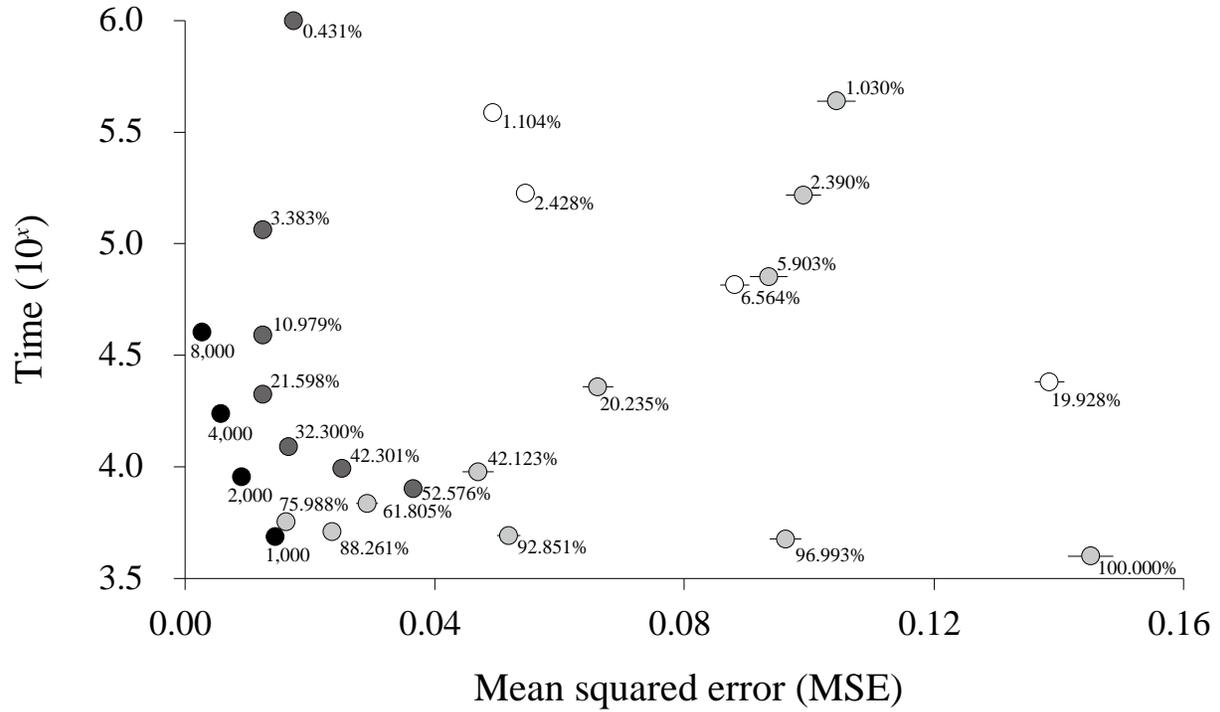

**Figure 4**. Comparisons of kernel-ABC and conventional ABCs under the realistic model using $s_{SFS}$.

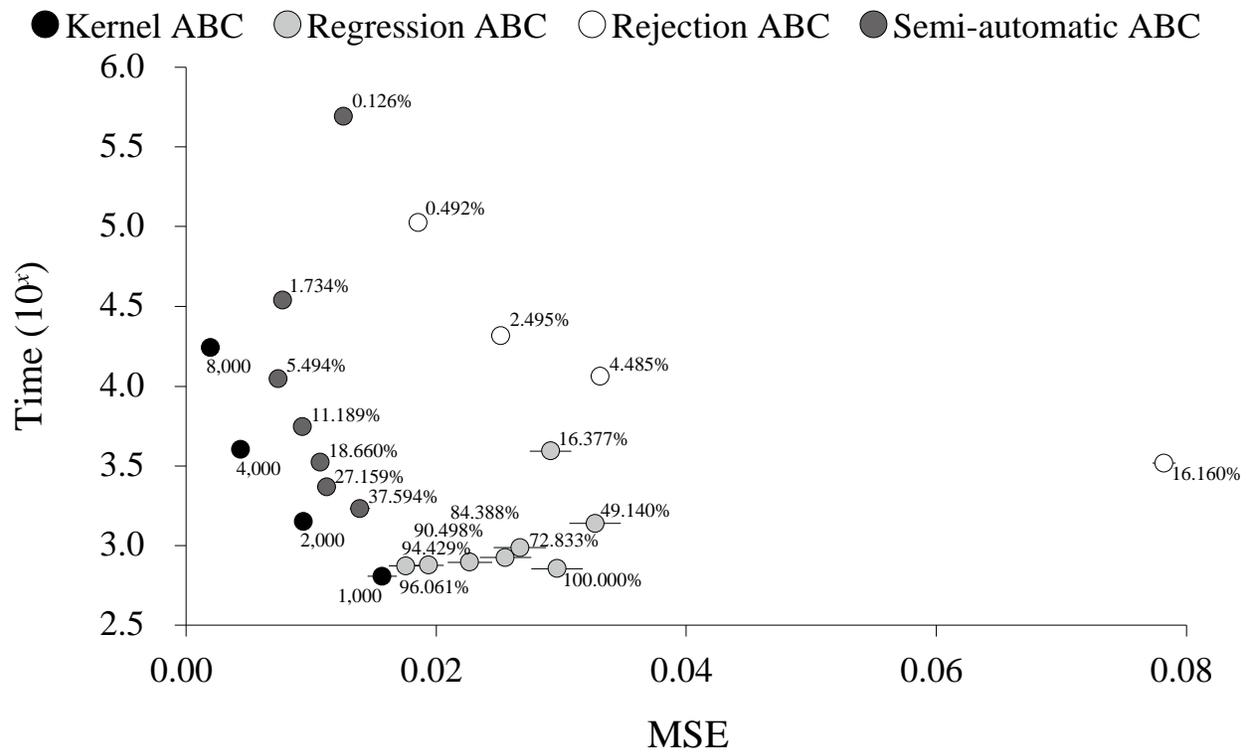

**Figure 5**. Comparisons of kernel-ABC and conventional ABCs under the realistic model using $s_{\text{SFS}} + s_{\text{HFS}}$.

1   **Supplementary Text**

2   **Example**. Suppose a simple beta-binomial model in which we assume a beta prior with

3   parameters $\alpha$. Assume we have a data that consists of $N$ trials, in which $M$ trials are

4   successes, and $N - M$ trials are failures. Suppose that $S$ does not contain any information,

5   whereas $T$ is the number of successes in the first $n$ trials. We have the posterior

6   distribution, $\pi(\theta|\mathcal{D}) \sim Beta(M + \alpha, N - M + \alpha)$, and the conditional distributions

7   $\pi(\theta|S) \sim Beta(\alpha, \alpha)$ and $\pi(\theta|T = t) \sim Beta(t + \alpha, n - t + \alpha)$. Then, the sufficient condition

8   for the decrease in the Kullback-Leibler divergence given by the proposition reduces to

$$\frac{N + 2\alpha - n - 1}{N + 2\alpha - 1} \frac{\binom{M + \alpha - 1}{t}\binom{N - M + \alpha - 1}{n - t}}{\binom{N + 2\alpha - 2}{n}} \geq \frac{\binom{-\alpha}{t}\binom{-\alpha}{n - t}}{\binom{-2\alpha}{n}}.$$

9   For simplicity, suppose $N, M \to \infty$ with $M/N \to p$. The left side is asymptotically the

10  probability mass function of the hypergeometric distribution, which tends to that of the

11  binomial distribution of the length, $n$, and the parameter, $p$. Therefore, the condition is

12  simplified and we have

$$\binom{n}{t} p^t (1-p)^{n-t} \geq \frac{\binom{-\alpha}{t}\binom{-\alpha}{n-t}}{\binom{-2\alpha}{n}}.$$

13  It is straightforward to see that the Kullback-Leibler divergence decreases if and only if

14  this condition is satisfied. In Supplementary Table, we present examples of outcomes that

15  decrease the Kullback-Leibler divergence for the case of $p = 1/2$. The Kullback-Leibler

16  divergence decreases if the conditional density given $T$ has the mode $(t + \alpha)/(n + 2\alpha)$

17  around the mode of the true posterior density given data, $p$. ∎

|     | $\alpha = 1$ |        | $\alpha = 11$ |        |
|-----|--------------|--------|---------------|--------|
| $n$ | $t$          | Prob.  | $t$           | Prob.  |
| 2   | 1            | 0.5000 | 1             | 0.5000 |
| 10  | 3 – 7        | 0.8906 | 4 – 6         | 0.6563 |
| 20  | 7 – 13       | 0.8847 | 8 – 12        | 0.7368 |
| 50  | 19 – 31      | 0.9351 | 21 – 29       | 0.7974 |
| 100 | 40 – 60      | 0.9648 | 43 – 57       | 0.8668 |
| 200 | 85 – 115     | 0.9718 | 89 – 111      | 0.8964 |

**Supplementary Table**: The outcomes and their probabilities that the Kullback-Leibler divergence decreases by an observation that first $n$ trials give $t$ successes. See Example in the text.